\documentclass[graybox]{svmult}

\usepackage{mathptmx}       
\usepackage{helvet}         
\usepackage{courier}        
\usepackage{type1cm}        
\usepackage{makeidx}         
\usepackage{graphicx}        
\usepackage{multicol}        
\usepackage[bottom]{footmisc}
\def\uu{\langle \bar u u \rangle}
\def\dd{\langle \bar d d \rangle}
\def\ss{\langle \bar s s \rangle}

\makeindex             

\begin{document}

\title*{Phenomenology  of the Heavy  Flavored spin 3/2 Baryons in Light Cone QCD }
\author{T. M. Aliev, K. Azizi, A. Ozpineci}
\institute{T. M. Aliev, \email{taliev@metu.edu.tr}
\and K. Azizi  \email{e146342@metu.edu.tr}
\and A. Ozpineci \email{ozpineci@metu.edu.tr} \\ \small 
Department of Physics, Middle East Technical University, 06531, Ankara, Turkey}

%
%
\maketitle

\abstract{Motivated by the results of the recent experimental discoveries for charm and bottom baryons,
  the masses and magnetic moments of the heavy  baryons with $J^P=3/2^+$ containing a single heavy quark are studied within  light cone QCD
  sum rules method. Our results on the masses of heavy baryons are in good agreement with predictions of other approaches, as well as with the
   existing experimental data.}

\section{Introduction}\label{sec:1}
In the recent years, considerable experimental progress has been made in the spectroscopy of baryons containing a single heavy quark. The CDF Collaboration
 has observed four bottom baryons $\Sigma^{\pm}_{b}$ and $\Sigma^{\ast\pm}_{b}$ \cite{Aaltonen1}. The DO \cite{Abazov} and CDF \cite{Aaltonen2} Collaborations have  seen the  $\Xi_{b}$. The BaBar
Collaboration discovered the $\Omega^{\ast}_{c}$ state
\cite{Aubert}. The CDF sensitivity appears adequate to observe new heavy baryons.
Study of the electromagnetic properties of baryons can give
noteworthy information on their internal structure. One of the main
static electromagnetic parameters of the baryons is their magnetic
moments. Magnetic moments of the heavy baryons in the framework of
different approaches are widely discussed in the literature. In the present work, we study the magnetic moments and masses of the
ground state baryons with total angular momentum 3/2 and containing
one heavy quark within light cone QCD sum rules. The paper is organized as follows. In section 2,
the light cone QCD sum rules for mass and magnetic moments of heavy baryons
are calculated. Section 3 is devoted to the numerical analysis of the mass and magnetic moment sum rules and
discussion. Detailed analysis of the mass and magnetic moments of the  baryons containing single heavy quark is presented in the original work in \cite{guzel}.
\section{Light cone QCD sum rules for the mass and magnetic moments of the heavy flavored  baryons}\label{sec:2}
 To calculate  the magnetic moments  of the heavy flavored hadrons, we start considering 
 the correlation function which is the  basic object in LCSR method. In this correlator, the  hadrons are represented by their interpolating quark currents. 
\begin{equation}\label{T}
T_{\mu\nu}=i\int d^{4}xe^{ipx}\langle0\mid T\{\eta_{\mu}(x)\bar{\eta}_{\nu}(0) \}\mid0\rangle_{\gamma},
\end{equation}
where $\eta_{\mu}$ is the interpolating current of the heavy baryon
and $\gamma$ means the external electromagnetic field. In QCD sum rules method, this
correlation function is calculated in two different ways: 1) In terms of quark-gluon language (QCD side), 2) In terms of hadrons, where the correlator is saturated by a tower of hadrons with the same quantum numbers as their interpolating currents (phenomenological side). The magnetic moments are
determined by matching two different representations of the
correlation function, i.e., theoretical and phenomenological forms,
using the dispersion relations.

From Eq. (\ref{T}), it follows that to calculate the correlation function from QCD side, we need the explicit expressions of the interpolating currents of heavy baryons with the angular momentum $J^P=3/2^+$. The main condition for constructing the interpolating currents from quark field is that they should have the same quantum numbers of the baryons under consideration. For the heavy baryons with $J^P=3/2^+$, the interpolating current is chosen in the following general form
\begin{eqnarray}\label{currentguy}
\eta_{\mu}=A\epsilon_{abc}\left\{\vphantom{\int_0^{x_2}}(q_{1}^{aT}C\gamma_{\mu}q_{2}^{b})Q^{c}+(q_{2}^{aT}C\gamma_{\mu}Q^{b})q_{1}^{c}+
(Q^{aT}C\gamma_{\mu}q_{1}^{b})q_{2}^{c}\right\},
\end{eqnarray}
where C is the charge conjugation operator and  a, b and c are color
indices. The value of A and quark fields $q_{1}$ and $q_{2}$  for
each heavy baryon is given in Table \ref{tab:1}.

\begin{table}
\caption{The value of A and quark fields $q_{1}$ and $q_{2}$  for
the corresponding baryons.}
\label{tab:1}      
\begin{tabular}{p{2cm}p{2cm}p{2cm}p{1.9cm}}
\hline\noalign{\smallskip}
&A & $q_{1}$ & $q_{2}$ \\
\noalign{\smallskip}\svhline\noalign{\smallskip}
$\Sigma_{b(c)}^{*+(++)}$&$1/\sqrt{3}$ &u&u\\
 $\Sigma_{b(c)}^{*0(+)}$&$\sqrt{2/3}$&u&d\\
 $\Sigma_{b(c)}^{*-(0)}$&$1/\sqrt{3}$&d&d\\
 $\Xi_{b(c)}^{*0(+)}$&$\sqrt{2/3}$&s&u\\
 $\Xi_{b(c)}^{*-(0)}$&$\sqrt{2/3}$&s&d\\
$\Omega_{b(c)}^{*-(0)}$ &$1/\sqrt{3}$&s &s\\
\noalign{\smallskip}\hline\noalign{\smallskip}
\end{tabular}
\end{table}
The phenomenological part of the correlation function can be obtained by inserting the complete set of states between the interpolating currents in (\ref{T}) with quantum numbers of heavy baryons.
\begin{eqnarray}\label{T2}
T_{\mu\nu}&=&\frac{\langle0\mid \eta_{\mu}\mid
B(p_{2})\rangle}{p_{2}^{2}-m_{B}^{2}}\langle B(p_{2})\mid
B(p_{1})\rangle_\gamma\frac{\langle B(p_{1})\mid
\bar{\eta}_{\nu}\mid 0\rangle}{p_{1}^{2}-m_{B}^{2}},
\end{eqnarray}
where $p_{1}=p+q$,  $p_{2}=p$ and q is the photon momentum. The vacuum to baryon matrix element of the interpolating current is defined as
\begin{equation}\label{lambdabey}
\langle0\mid \eta_{\mu}(0)\mid B(p,s)\rangle=\lambda_{B}u_{\mu}(p,s),
\end{equation}
where $\lambda_{B}$ is the  residue and $u_{\mu}(p,s)$ is the Rarita-Schwinger spinor. The matrix element $\langle B(p_{2})\mid
B(p_{1})\rangle_\gamma$ entering Eq.  (\ref{T2}) can be parameterized in terms of the form factors $f_{i}$ and $G_{i}$ as follows
\begin{eqnarray}\label{matelpar}
\langle B(p_{2})\mid
B(p_{1})\rangle_\gamma &=&\varepsilon_{\rho}\bar u_{\mu}(p_{2})\left\{-g_{\mu\nu}\left[\vphantom{\int_0^{x_2}}\gamma_{\rho}(f_{1}+f_{2})\right.
+\frac{(p_{1}+p_{2})_{\rho}}{2m_{B}}f_{2}+q_{\rho}f_{3}\right]
\nonumber\\&-&\frac{q_{\mu}q_{\nu}}{(2m_{B})^{2}}\left[\gamma_{\rho}(G_{1}+G_{2})+
\frac{(p_{1}+p_{2})_{\rho}}{2m_{B}}G_{2}+q_{\rho}G_{3}\right]\left.\vphantom{\int_0^{x_2}}\right\}\bar u_{\nu}(p_{1}),\nonumber\\
\end{eqnarray}
where $\varepsilon_{\rho}$ is the photon polarization vector and $q^2=(p_{1}-p_{2})^2$. To obtain the explicit expressions of the correlation function,  summation over spins of the spin 3/2 particles is performed. 
Using the above equations in principle one can write down the phenomenological part of the correlator. But, the following two drawbacks appear: a) all Lorentz structures are not independent, b) not only spin 3/2, but spin 1/2 states  also contribute to the correlation function. Indeed the matrix element of the current $\eta_{\mu}$ between vacuum and spin 1/2 states is nonzero and is determined as
\begin{equation}\label{spin12}
\langle0\mid \eta_{\mu}(0)\mid B(p,s=1/2)\rangle=\alpha (4  p_{\mu}- m\gamma_{\mu})u(p,s=1/2),
\end{equation}
where the condition $\gamma_\mu \eta^\mu = 0$ is imposed.

There are two different ways to remove the unwanted spin 1/2 contribution and retain only independent structures in the
correlation function: 1) Introduce projection operators for the spin 3/2 states, which kill the spin 1/2 contribution,
2) Ordering Dirac matrices  in a specific order and eliminate the structures that receive contributions from spin $1/2$ states.
In this work, we will follow the second method and choose the  ordering for Dirac
matrices as $\gamma_{\mu}\not\!p\not\!\varepsilon\not\!q\gamma_{\nu}$. With this ordering for the correlator, we get
\begin{eqnarray}\label{final phenpart}
T_{\mu\nu}&=&\lambda_{B}^{2}\frac{1}{(p_{1}^{2}-m_{B}^{2})(p_{2}^{2}-m_{B}^{2})}
\left[g_{\mu\nu}\not\!p\not\!\varepsilon\not\!q\frac{g_{M}}{3}\right.\nonumber\\&+&\mbox{other structures with }\gamma_{\mu}\mbox{at the beginning and} \gamma_{\nu}\mbox{ at the end } \nonumber\\
&&\mbox{  or
which are proportional to }p_{2\mu}\mbox{or }p_{1\nu}\left.\vphantom{\int_0^{x_2}}\right],
\end{eqnarray}
where $g_{M}/3=f_{1}+f_{2}$ and at $q^2=0$, $g_{M}$ is the magnetic moment of the baryon in units of its natural
 magneton. The factor 3 is due the fact that in the non-relativistic limit the interaction Hamiltonian
 with magnetic field is equal to $g_M B = 3(f_{1}+f_{2})B$.
\begin{figure}[b]
\sidecaption
\includegraphics[scale=.20]{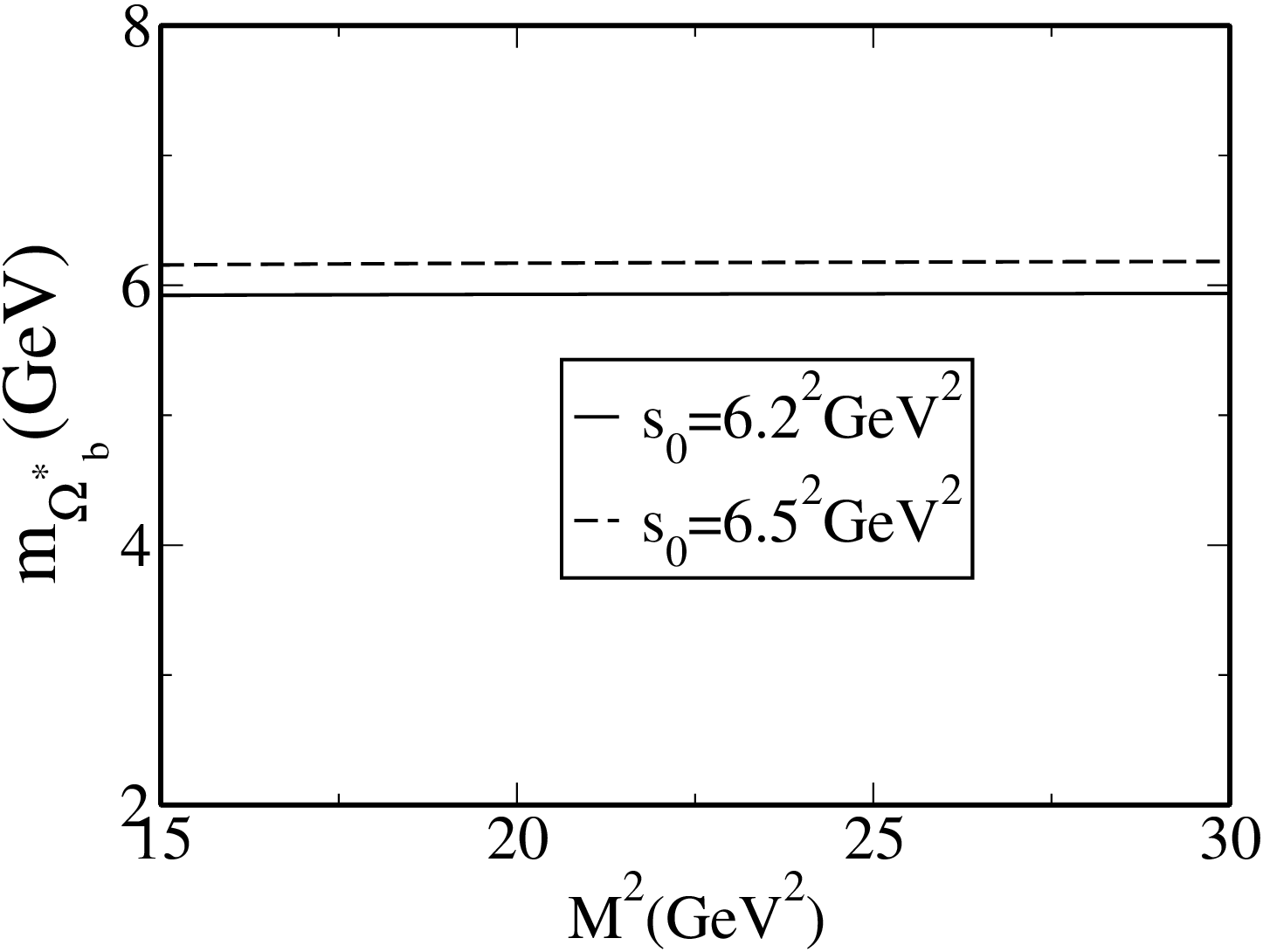}
\caption {The dependence of  mass of the $\Omega_{b}^{*}$ on the
Borel parameter $M^{2}$ for two  fixed values of continuum
threshold $s_{0}$.}
\label{fig:1}       
\end{figure}
\begin{figure}[b]
\sidecaption
\includegraphics[scale=.20]{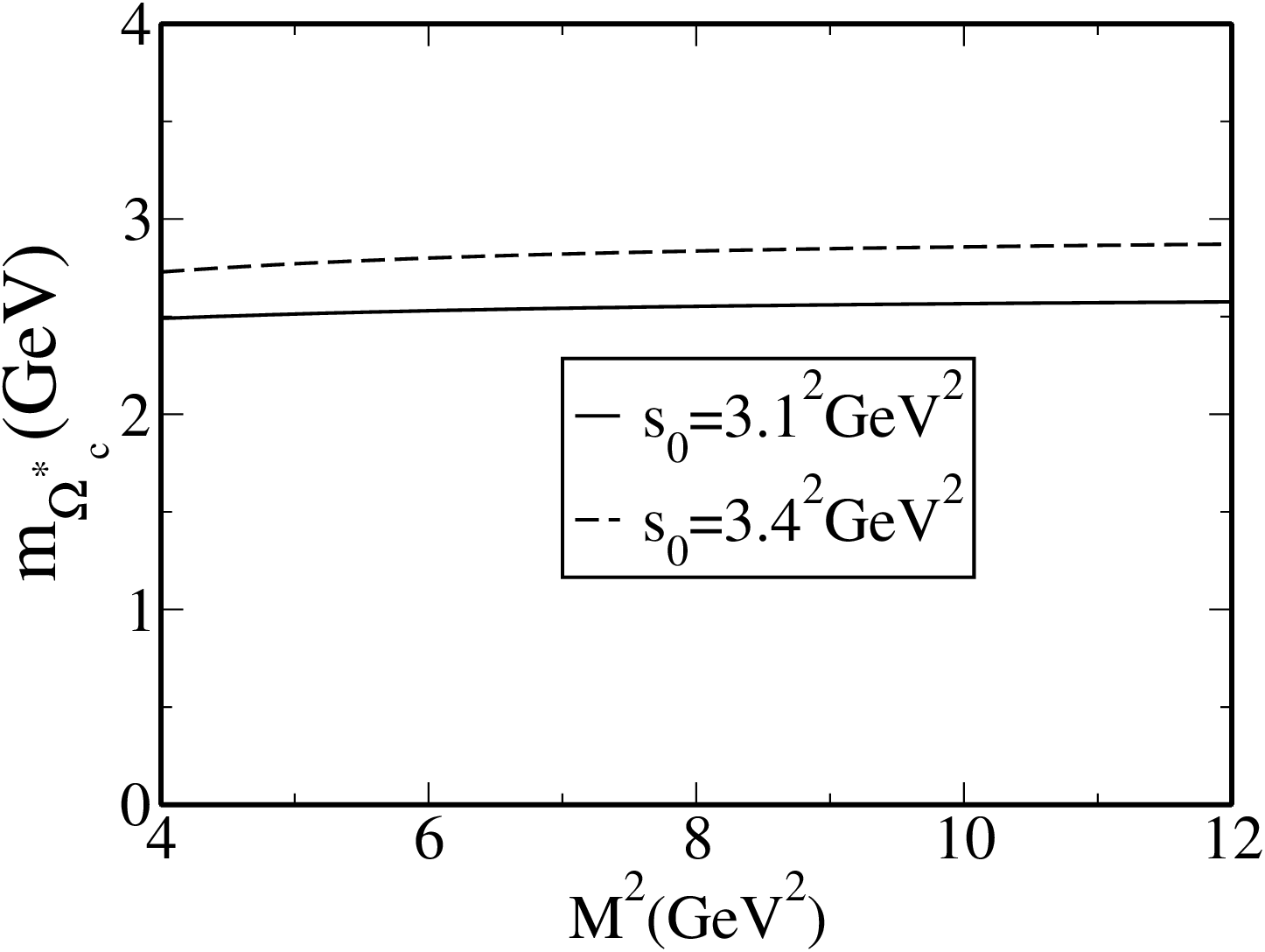}
\caption {The dependence of  mass of the $\Omega_{c}^{*}$ on the
Borel parameter $M^{2}$ for two  fixed values of continuum
threshold $s_{0}$.}
\label{fig:2}       
\end{figure}
\begin{figure}[b]
\sidecaption
\includegraphics[scale=.20]{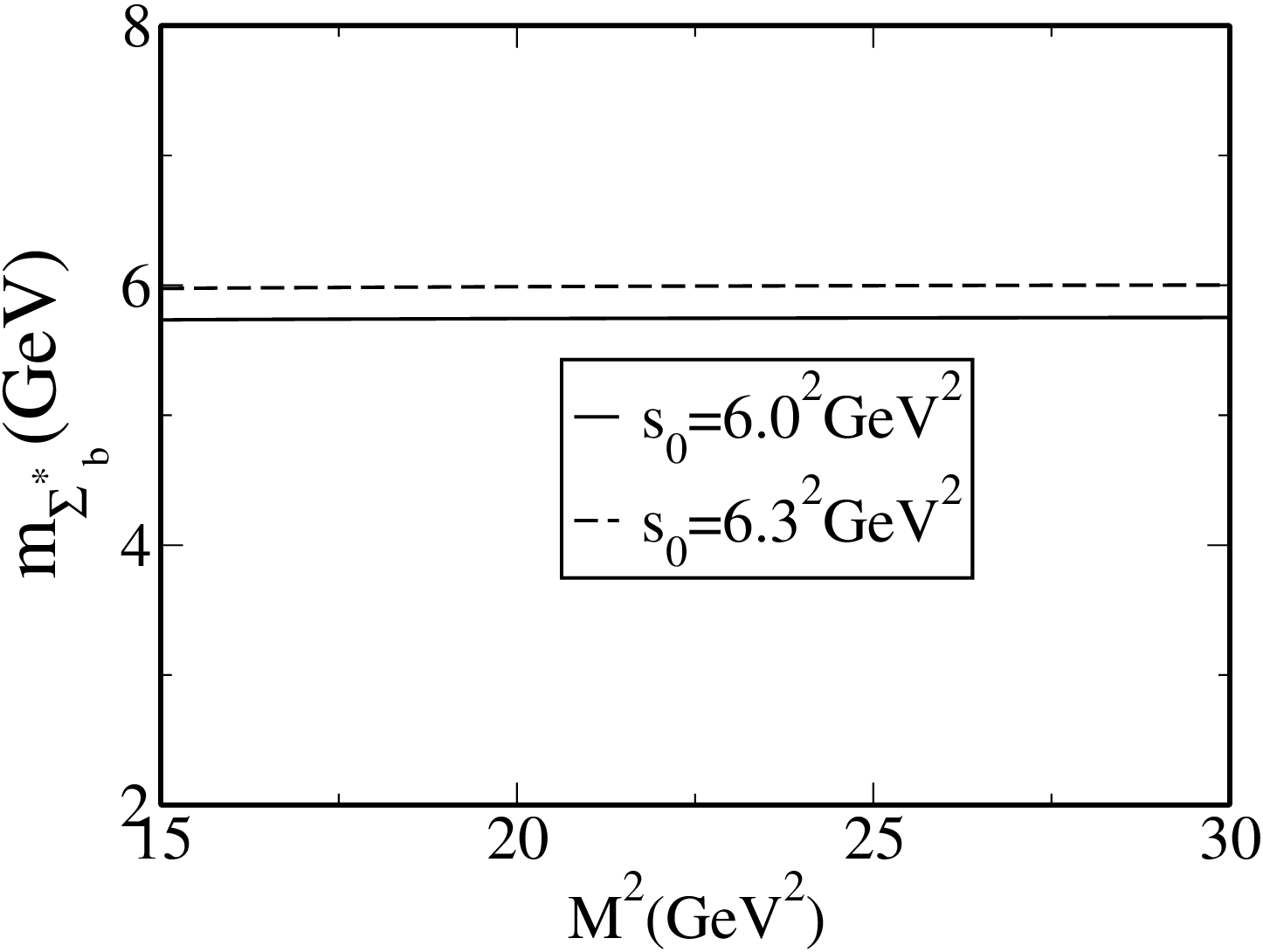}
\caption {The same as Fig. 1, but for $\Sigma_{b}^{*}$.}
\label{fig:3}       
\end{figure}
\begin{figure}[b]
\sidecaption
\includegraphics[scale=.20]{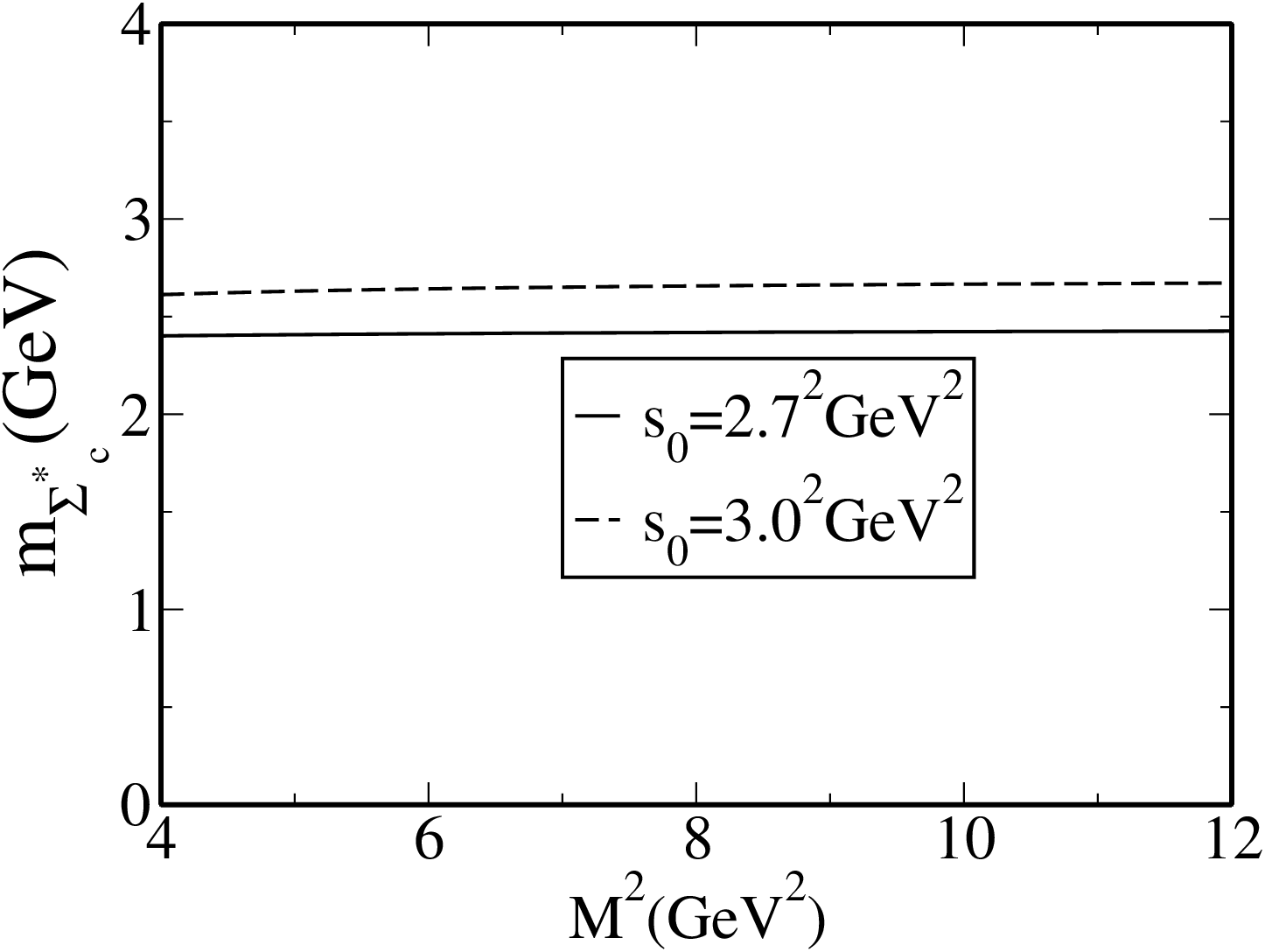}
\caption {The same as Fig. 2, but for  $\Sigma_{c}^{*}$.}
\label{fig:4}       
\end{figure}
\begin{figure}[b]
\sidecaption
\includegraphics[scale=.20]{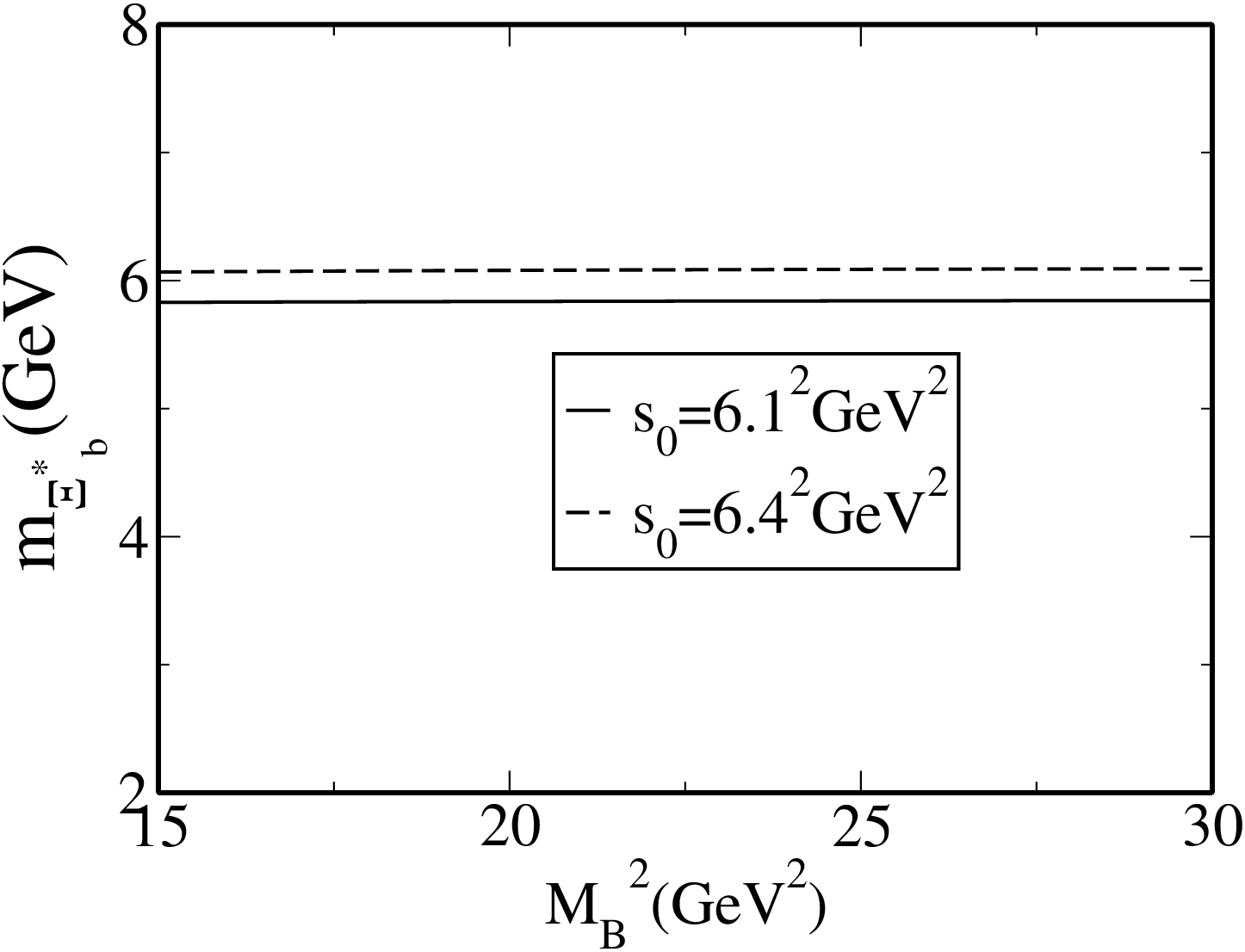}
\caption {The same as Fig. 1, but for $\Xi_{b}^{*}$.}
\label{fig:5}       
\end{figure}
\begin{figure}[b]
\sidecaption
\includegraphics[scale=.20]{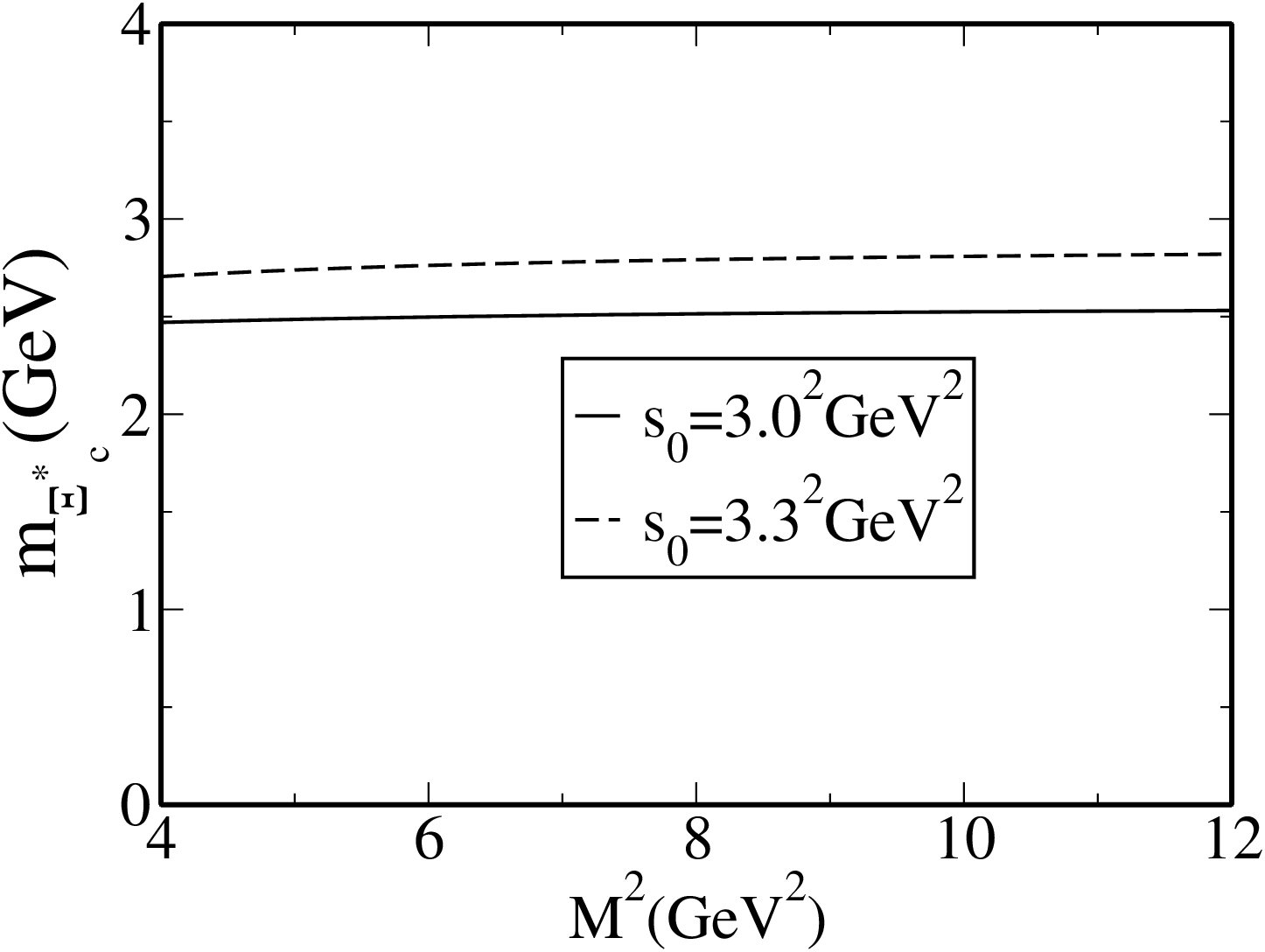}
\caption {The same as Fig. 2, but for$\Xi_{c}^{*}$.}
\label{fig:6}       
\end{figure}
\begin{figure}[b]
\sidecaption
\includegraphics[scale=.20]{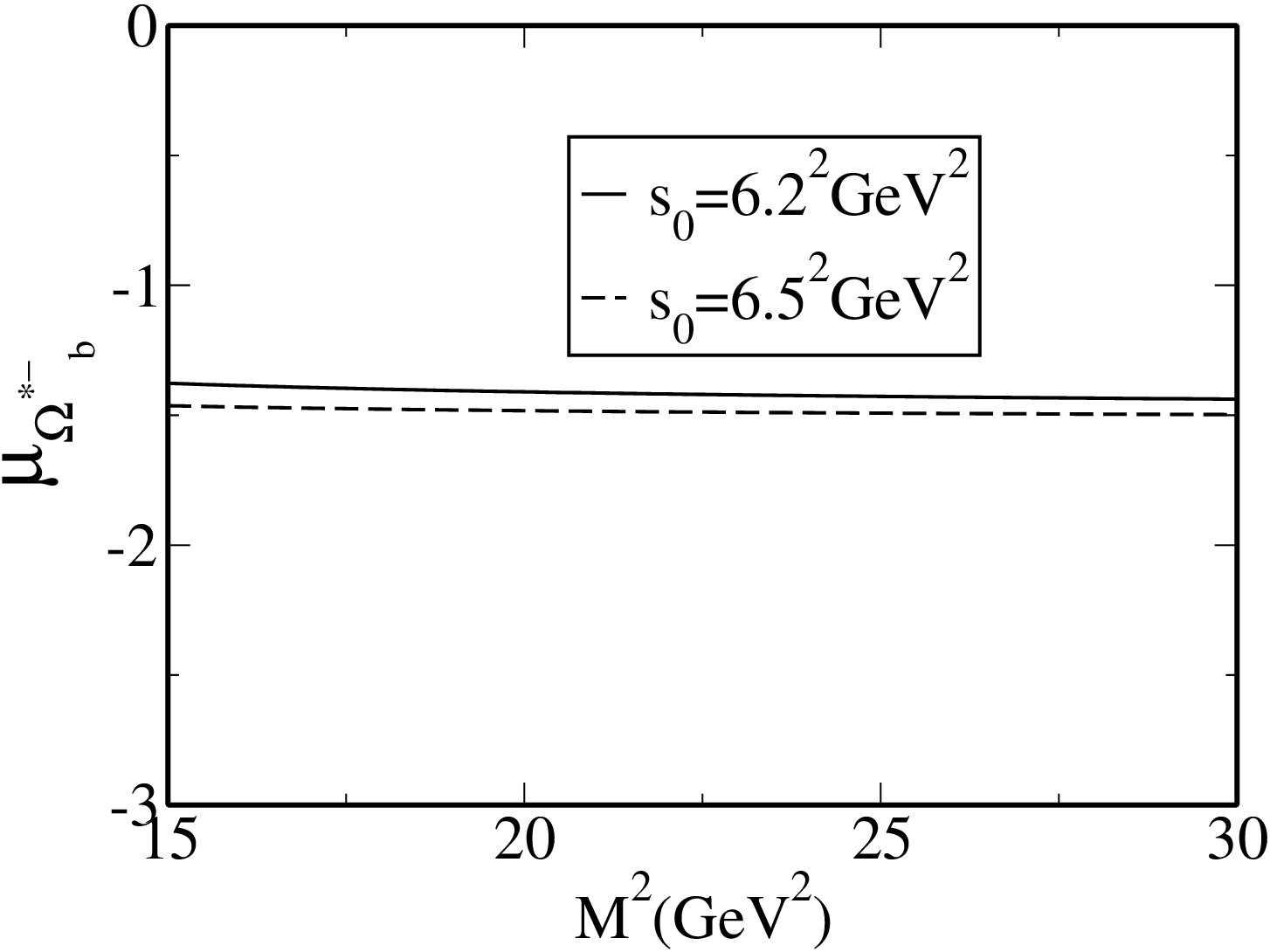}
\caption {The dependence of the magnetic moment of $\Omega_{b}^{*-}$ on Borel parameter
$M^{2}$ (in units of nucleon magneton) at two fixed values of $s_{0}$.}
\label{fig:7}       
\end{figure}
\begin{figure}[b]
\sidecaption
\includegraphics[scale=.20]{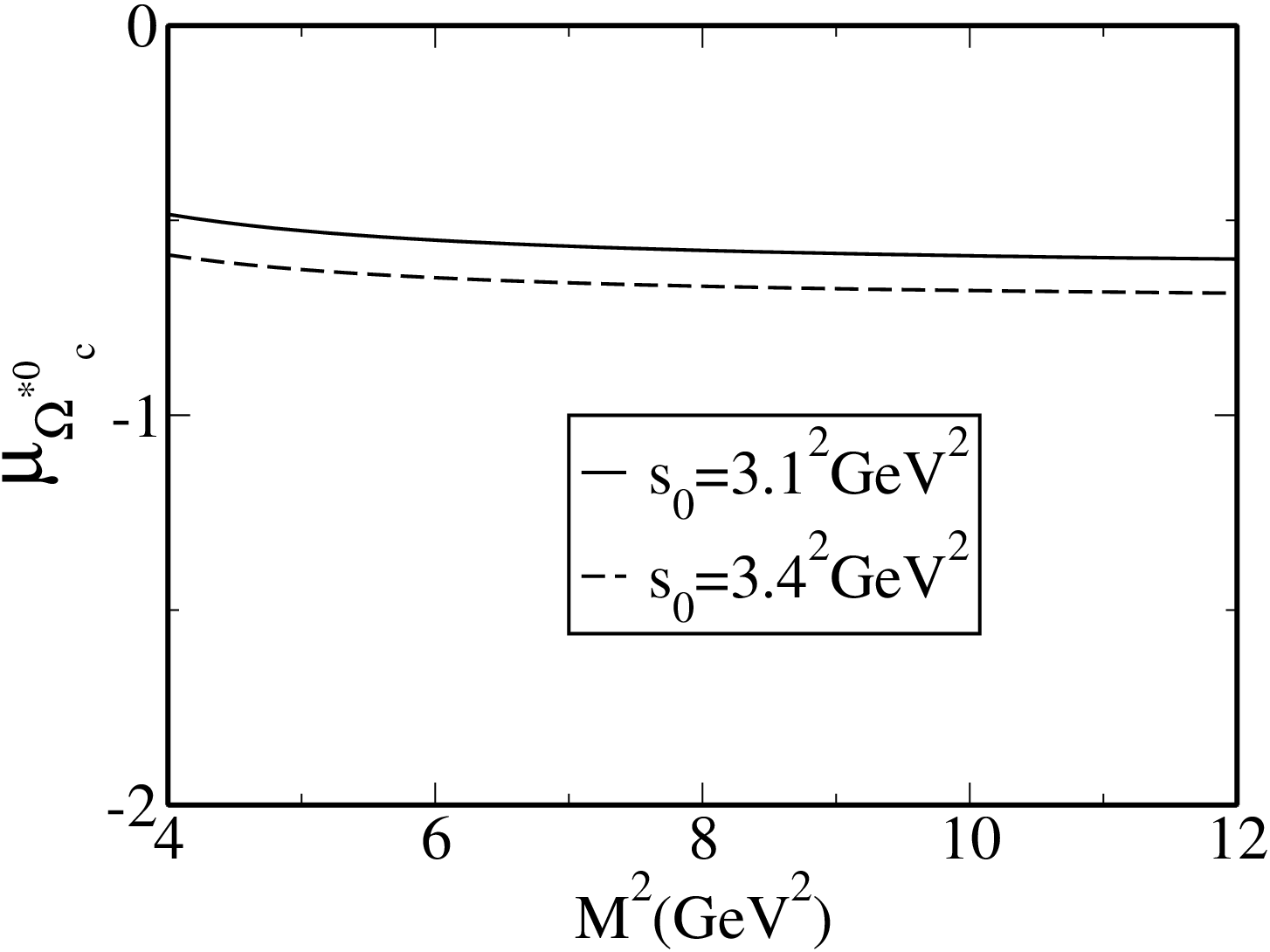}
\caption {The dependence of the magnetic moment of $\Omega_{c}^{*0}$ on Borel parameter
$M^{2}$ (in units of nucleon magneton) at two fixed values of $s_{0}$.}
\label{fig:8}       
\end{figure}
\begin{figure}[b]
\sidecaption
\includegraphics[scale=.20]{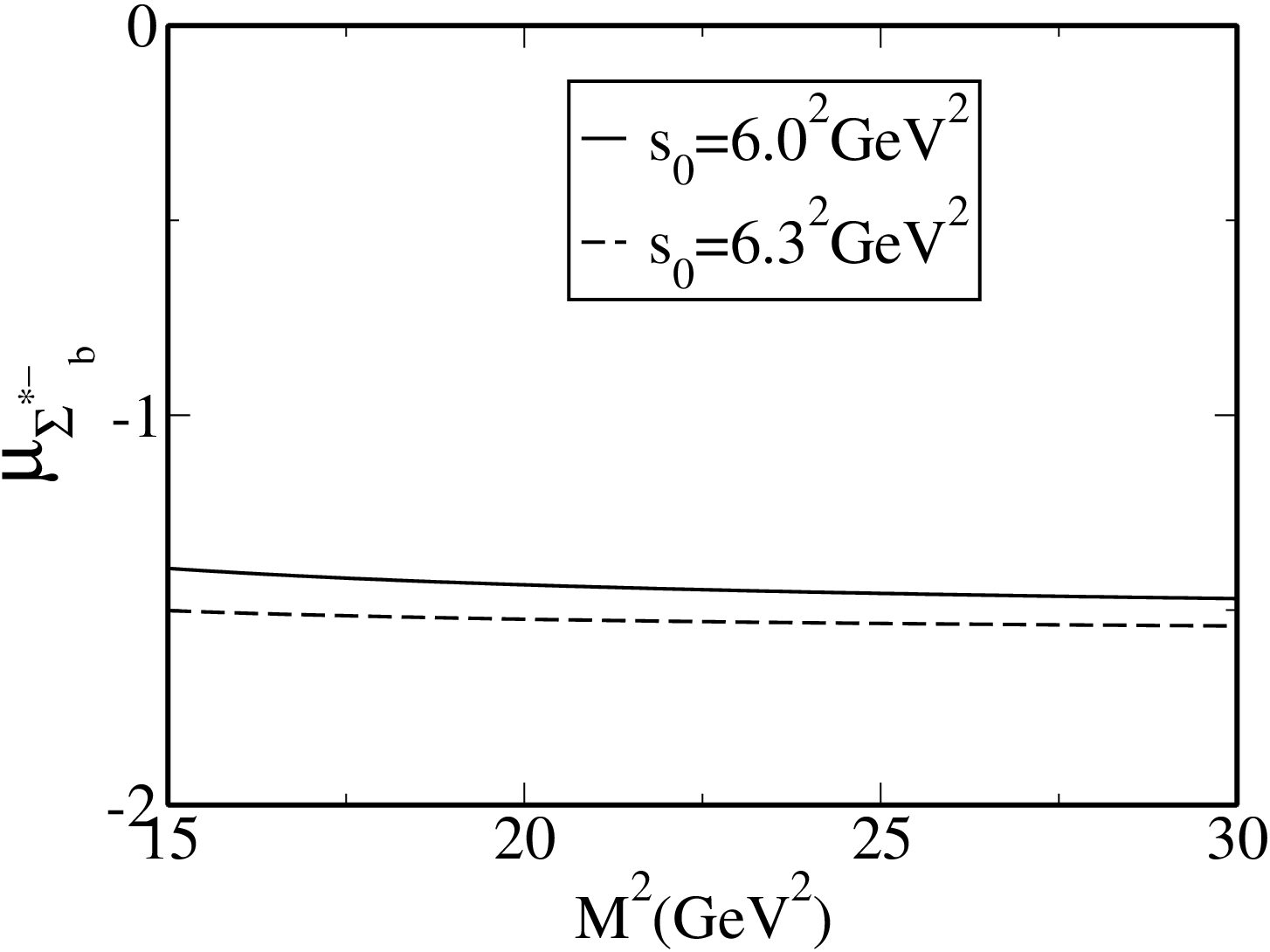}
\caption {The same as Fig. 7, but for  $\Sigma_{b}^{*-}$.}
\label{fig:9}       
\end{figure}
\begin{figure}[b]
\sidecaption
\includegraphics[scale=.20]{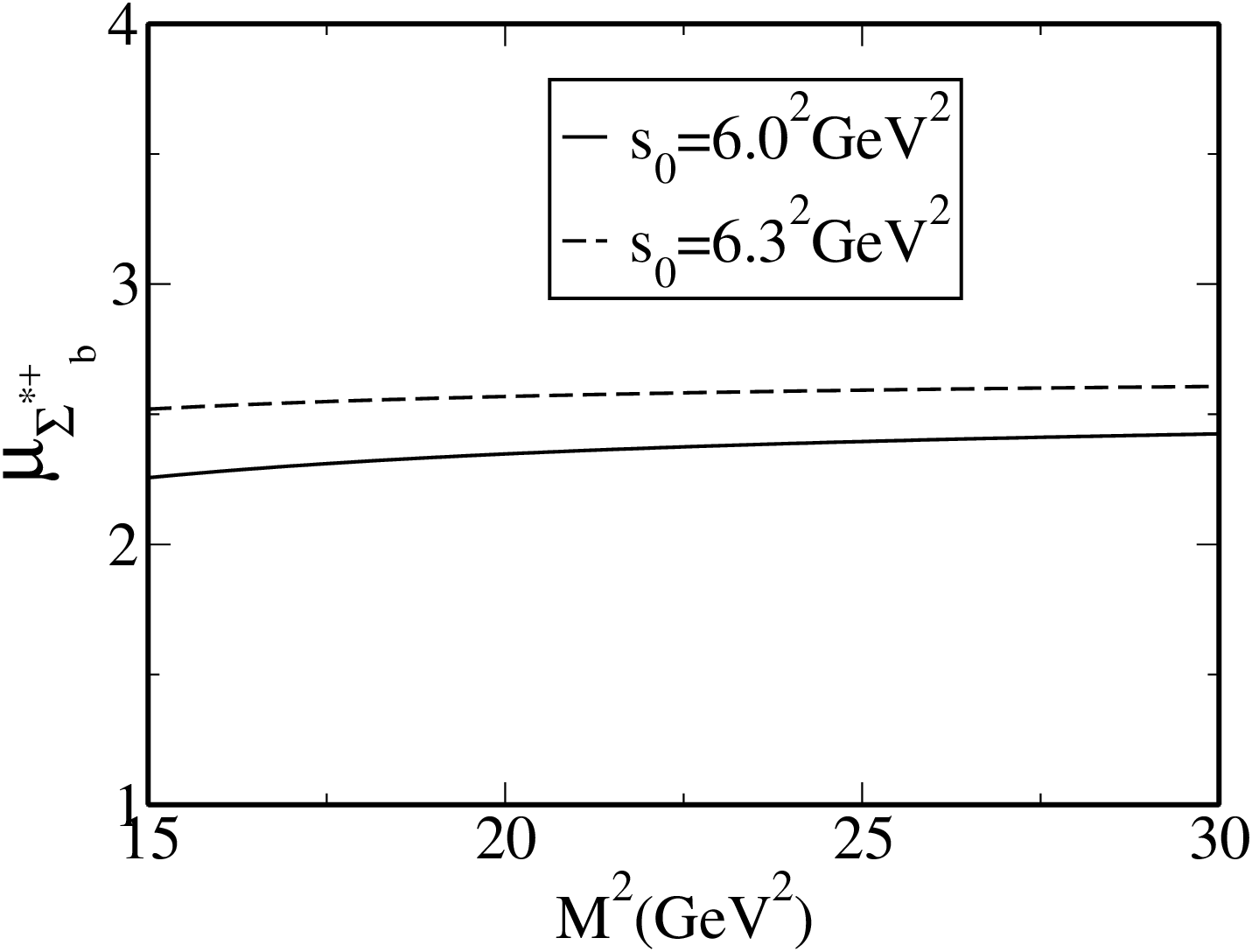}
\caption {The same as Fig. 7, but for  $\Sigma_{b}^{*+}$.}
\label{fig:10}       
\end{figure}
\begin{figure}[b]
\sidecaption
\includegraphics[scale=.20]{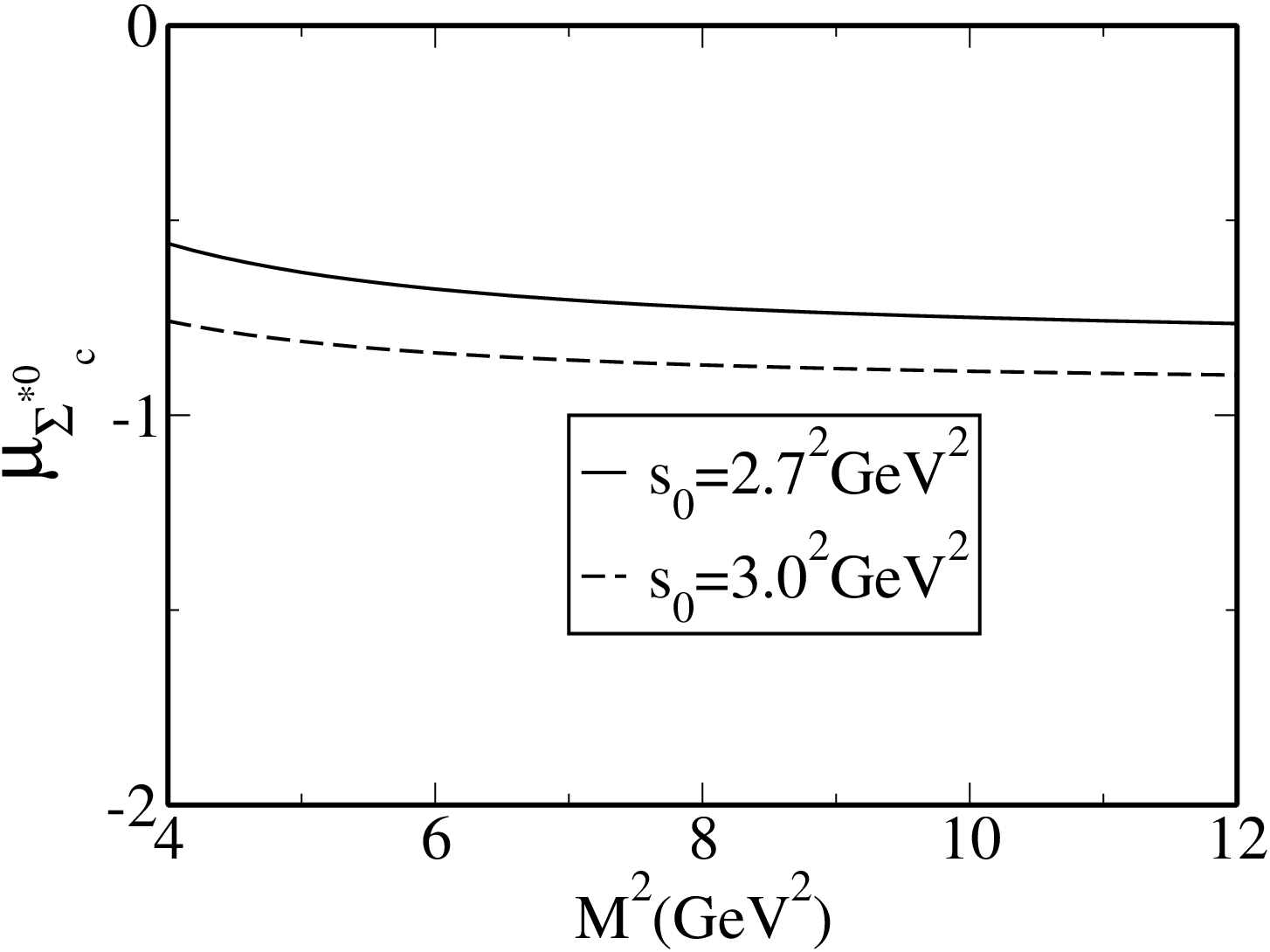}
\caption {The same as Fig. 8, but for  $\Sigma_{c}^{*0}$.}
\label{fig:11}       
\end{figure}
\begin{figure}[b]
\sidecaption
\includegraphics[scale=.20]{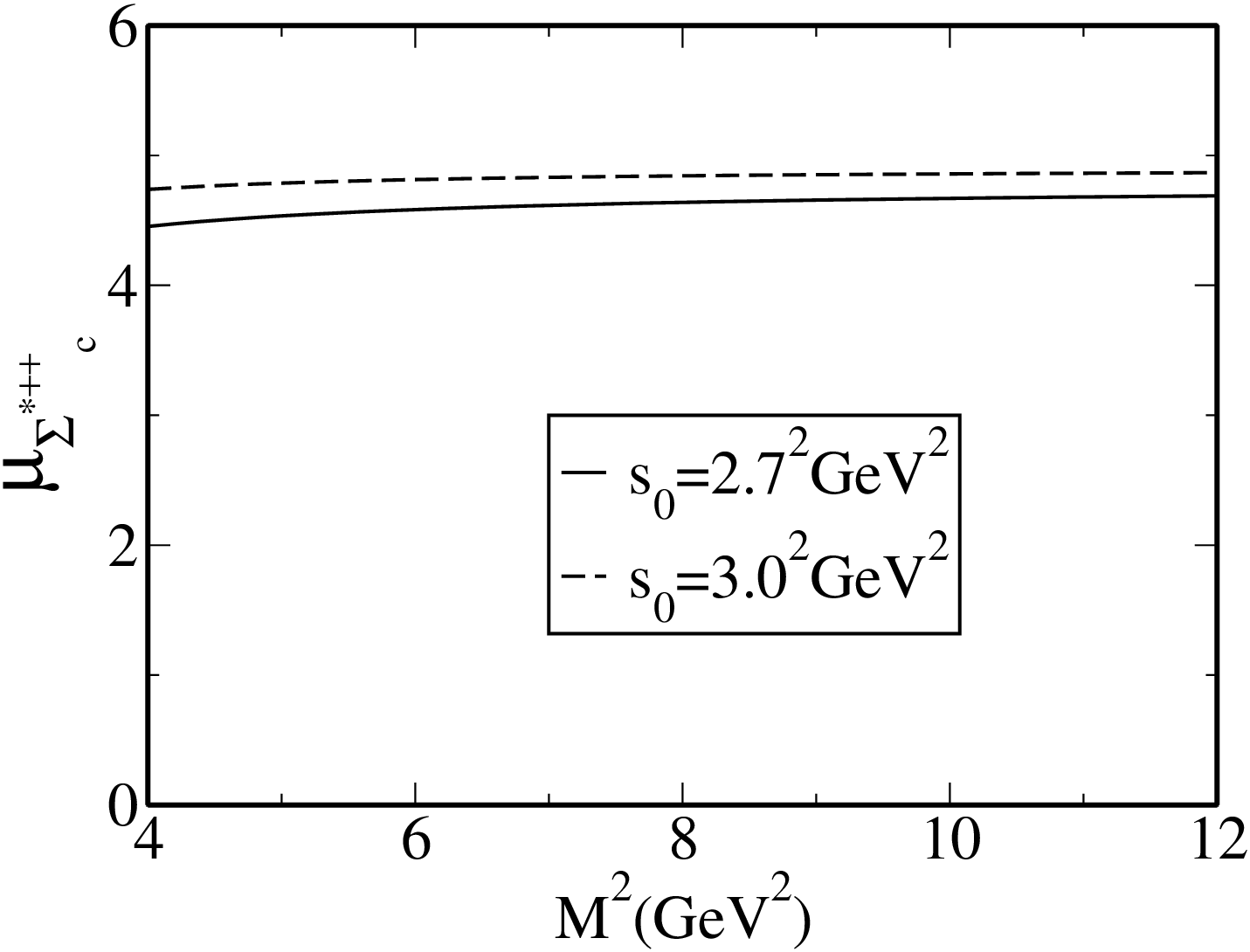}
\caption {The same as Fig. 8, but for $\Sigma_{c}^{*++}$.}
\label{fig:12}       
\end{figure}
\begin{figure}[b]
\sidecaption
\includegraphics[scale=.20]{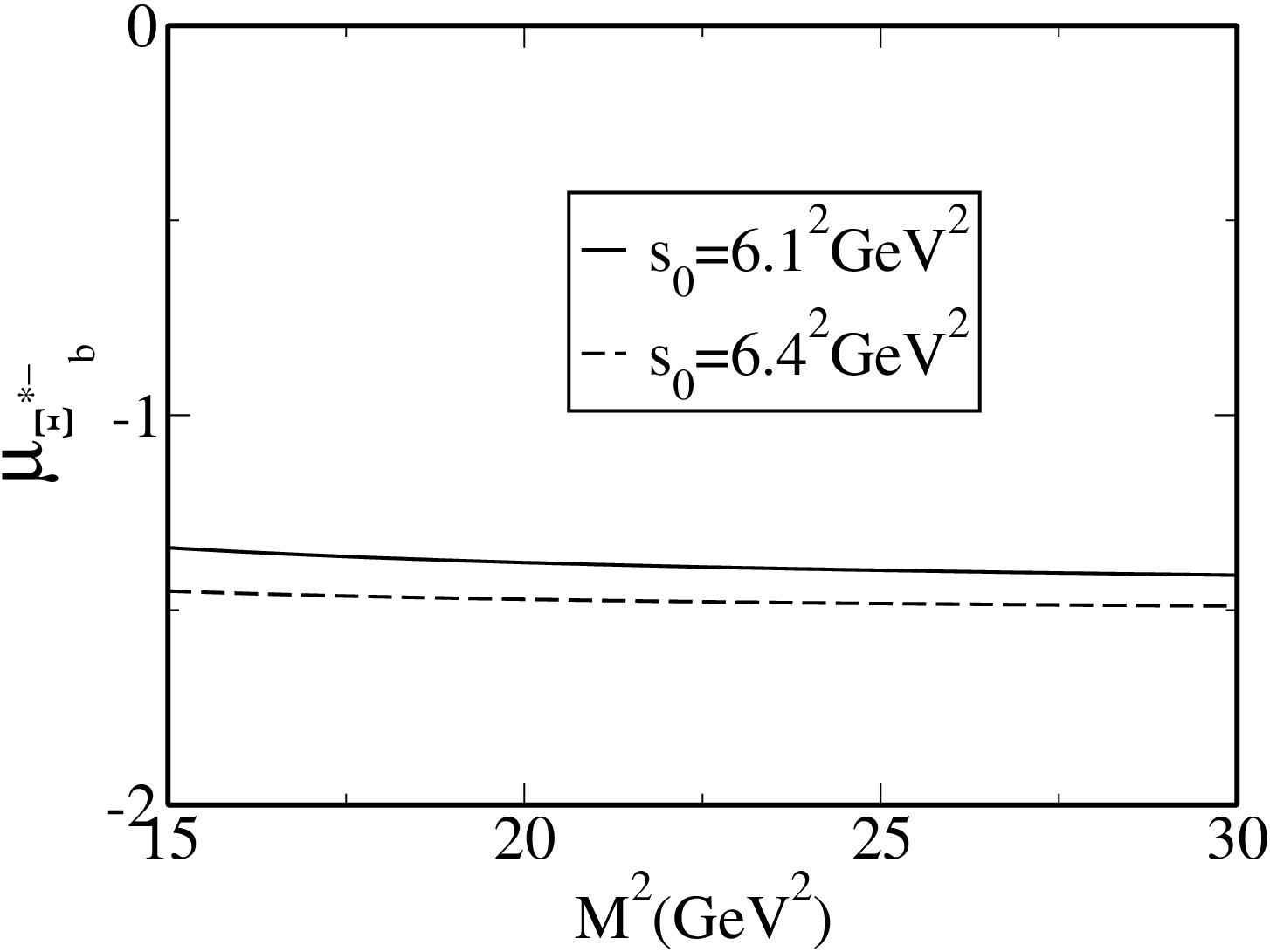}
\caption {The same as Fig. 7, but for  $\Xi_{b}^{*-}$.}
\label{fig:13}       
\end{figure}
\begin{figure}[b]
\sidecaption
\includegraphics[scale=.20]{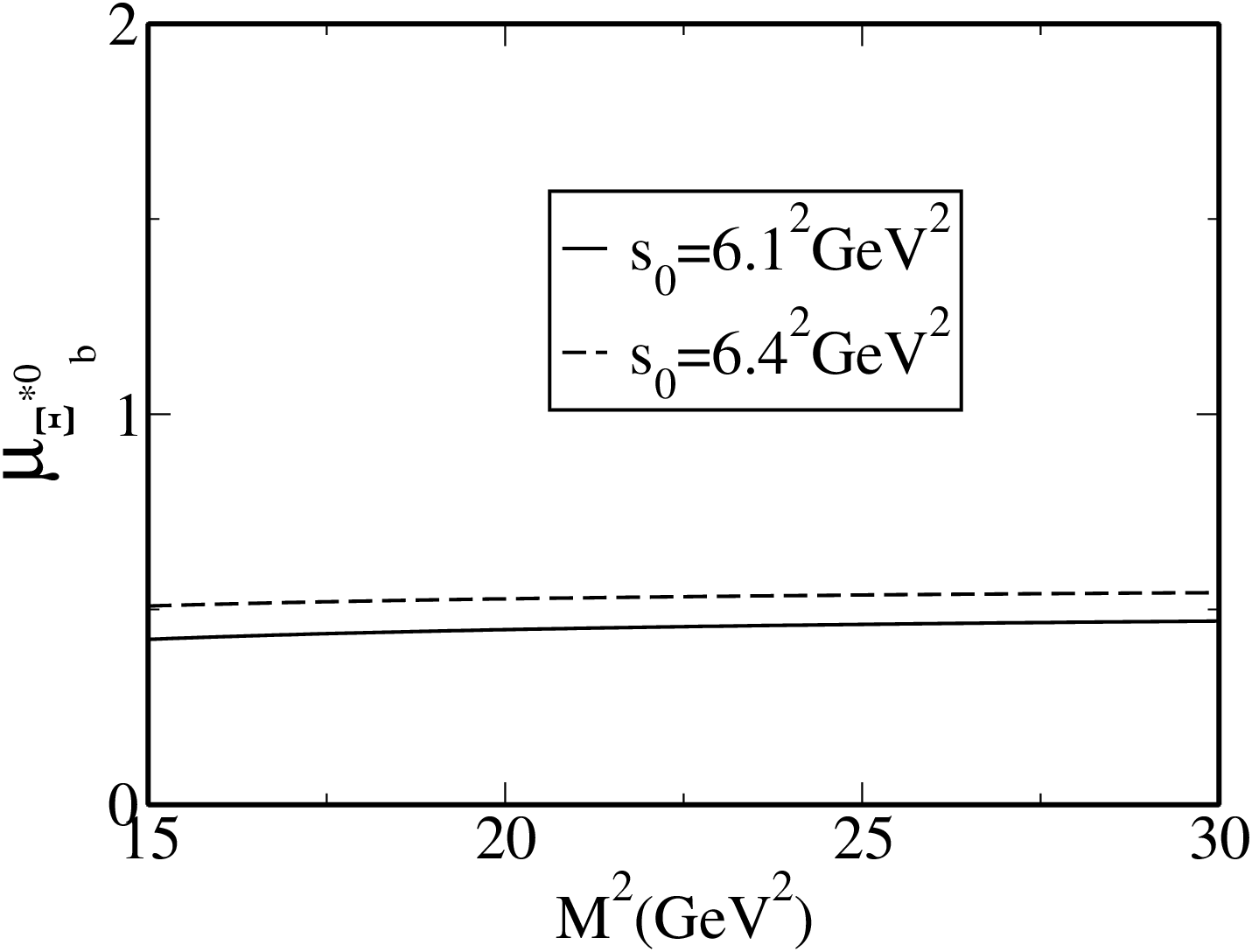}
\caption {The same as Fig. 7, but for  $\Xi_{b}^{*0}$.}
\label{fig:14}       
\end{figure}
\begin{figure}[b]
\sidecaption
\includegraphics[scale=.20]{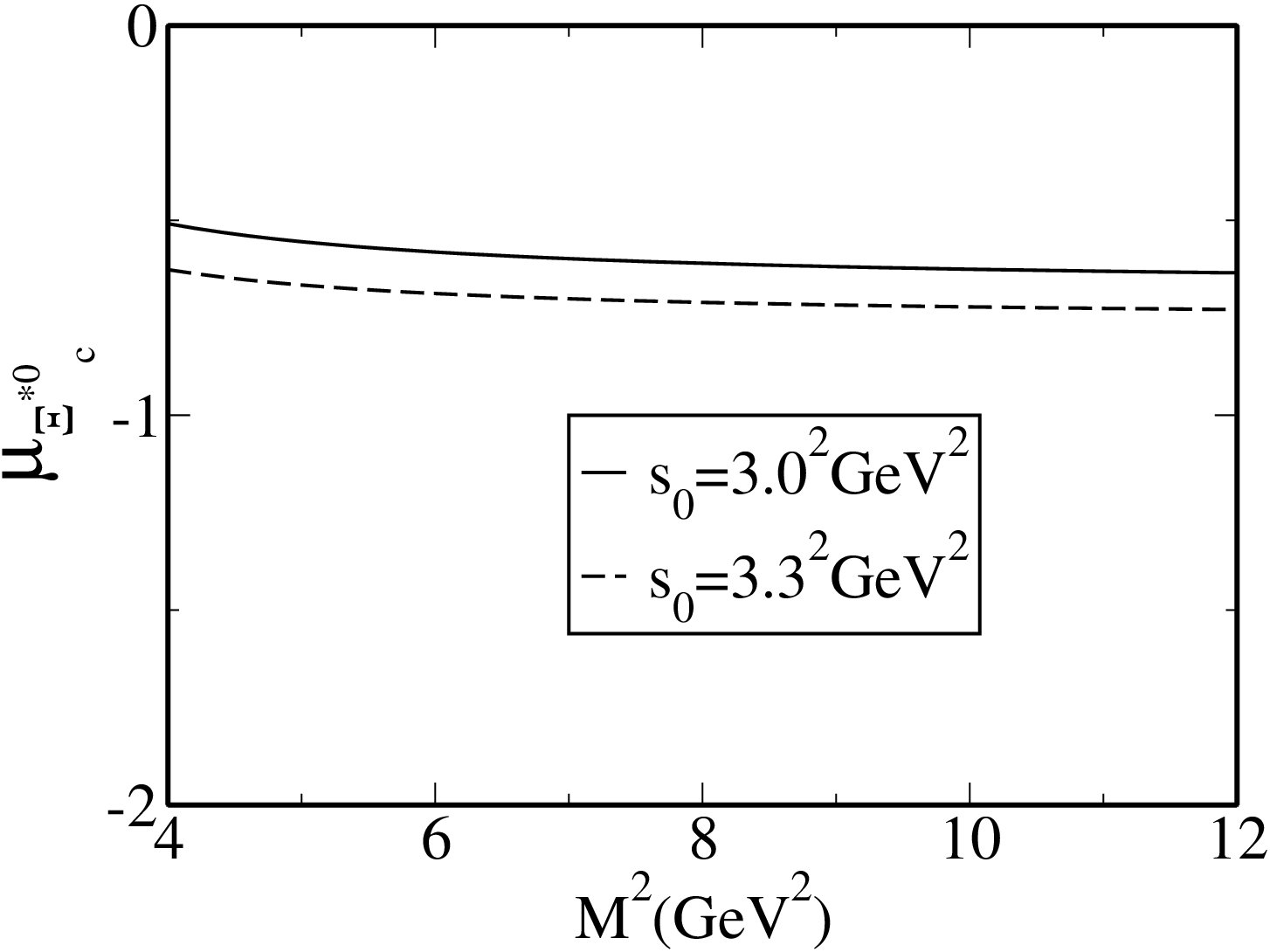}
\caption {The same as Fig. 8, but for  $\Xi_{c}^{*0}$.}
\label{fig:15}       
\end{figure}
\begin{figure}[b]
\sidecaption
\includegraphics[scale=.20]{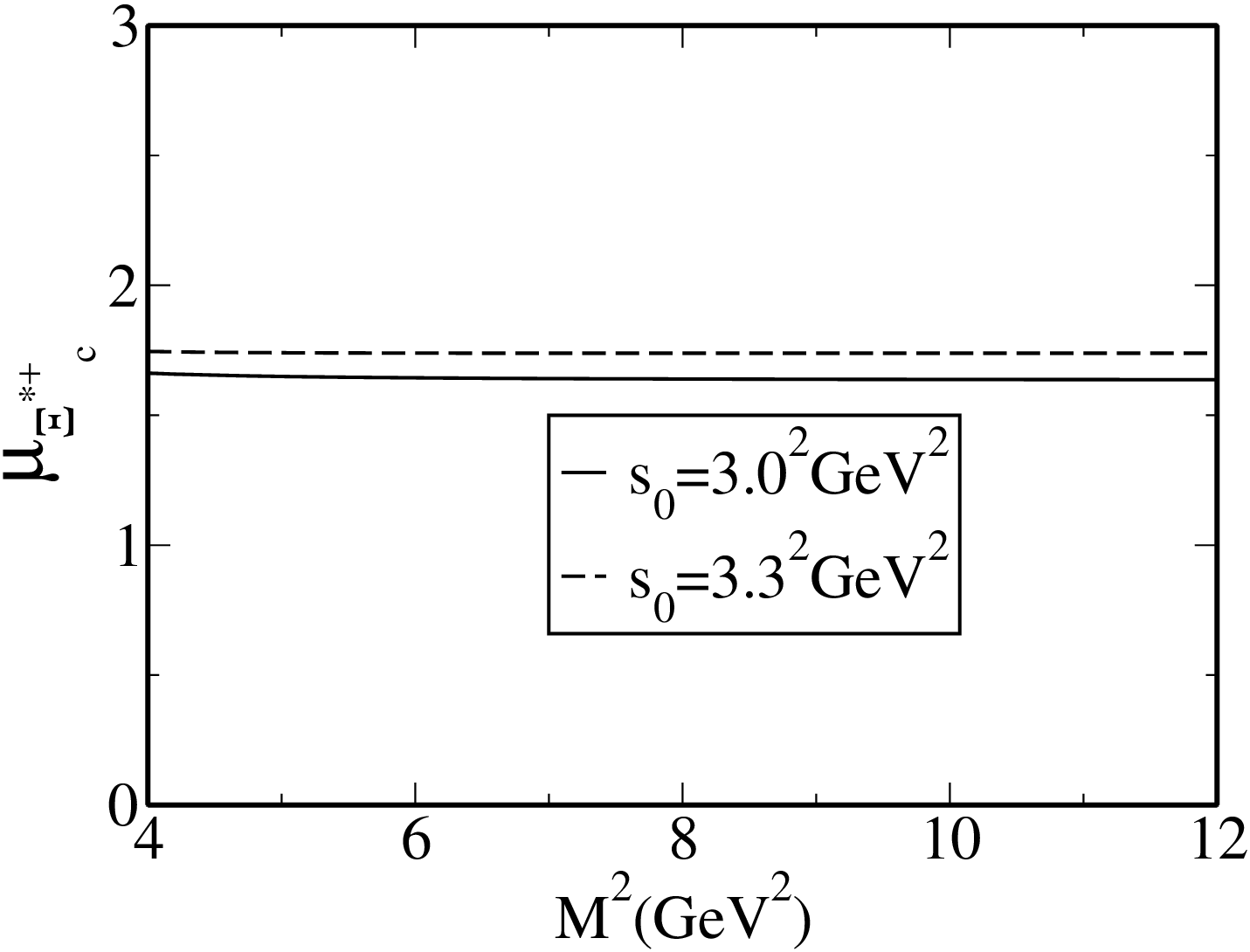}
\caption {The same as Fig. 8, but for  $\Xi_{c}^{*+}$.}
\label{fig:16}       
\end{figure}

On QCD side, the correlation function (\ref{T}) can be evaluated using operator product expansion.  After simple calculations, we get the following expression for the correlation
function in terms of quark propagators
\begin{eqnarray}\label{tree expresion.m}
\Pi_{\mu\nu}&=&-iA^{2}\epsilon_{abc}\epsilon_{a'b'c'}\int
d^{4}xe^{ipx}\langle0[\gamma(q)]\mid\{S_{Q}^{ca'}
\gamma_{\nu}S'^{bb'}_{q_{2}}\gamma_{\mu}S_{q_{1}}^{ac'}\nonumber\\&+&S_{Q}^{cb'}
\gamma_{\nu}S'^{aa'}_{q_{1}}\gamma_{\mu}S_{q_{2}}^{bc'}+ S_{q_{2}}^{ca'}
\gamma_{\nu}S'^{bb'}_{q_{1}}\gamma_{\mu}S_{Q}^{ac'}+S_{q_{2}}^{cb'}
\gamma_{\nu}S'^{aa'}_{Q}\gamma_{\mu}S_{q_{1}}^{bc'}\nonumber\\&+& S_{q_{1}}^{cb'}
\gamma_{\nu}S'^{aa'}_{q_{2}}\gamma_{\mu}S_{Q}^{bc'}+S_{q_{1}}^{ca'}
\gamma_{\nu}S'^{bb'}_{Q}\gamma_{\mu}S_{q_{2}}^{ac'}
+Tr(\gamma_{\mu}S_{q_{1}}^{ab'}\gamma_{\nu}S'^{ba'}_{q_{2}})S^{cc'}_{Q}\nonumber\\&+&Tr(\gamma_{\mu}S_{Q}^{ab'}\gamma_{\nu}S'^{ba'}_{q_{1}})S^{cc'}_{q_{2}}+Tr(\gamma_{\mu}S_{q_{2}}^{ab'}\gamma_{\nu}S'^{ba'}_{Q})S^{cc'}_{q_{1}}\}\mid 0\rangle,
\end{eqnarray}
where $S'=CS^TC$ and $S_{Q}(S_{q})$ is the full heavy (light) quark
propagator. In calculation of the correlation function from QCD
side, we take into account terms linear in $m_{q}$  and neglect
quadratic terms.The correlator contains three different
contributions: 1) Perturbative contributions, 2) Mixed
contributions, i.e., the photon is radiated from freely propagating
quarks at short distance and at least one of quark pairs interact
with QCD vacuum non-perturbatively. The last interaction is
parameterized in terms of quark condensates. 3) Non-perturbative
contributions, i.e., when photon is radiated at long distances. In order to calculate the contributions of the photon emission from large distances,  the matrix
 elements of nonlocal operators $\bar q \Gamma_{i}q$ between the photon
 and vacuum states are needed, $ \langle\gamma(q)\mid\bar q
 \Gamma_{i}q\mid0\rangle$. These matrix  elements are determined  in terms of
 the photon distribution amplitudes (DA's). For these matrix elements and also  the photon DA's see \cite{Ball}.

Using the expressions of the light and heavy full  propagators and the photon DA's and separating the coefficient of the structure
$g_{\mu\nu}\not\!p\not\!\varepsilon\not\!q$, the expression of the correlation function from QCD side is obtained. Separating the coefficient of the
same structure from phenomenological part and equating these representations of the correlator,  sum rules for
the magnetic moments of the $J^P=3/2^+$ heavy baryons is obtained. In order to suppress the contribution of higher states and
continuum, Borel transformation with respect to the variables $p_{2}^2=p^2$ and $p_{1}^2=(p+q)^2$
 is applied.The sum rules for the magnetic moments is obtained
as
\begin{eqnarray}\label{magneticmoment1}
-\lambda_{B}^{2}\frac{\mu_{B_{Q}}}{3}e^{\frac{-m_{B_{Q}}^{2}}{M^{2}}}=A^2\Pi^{B_Q}.
\end{eqnarray}
The functions $\Pi_i(q_1,q_2,Q)$ can be written as:
\begin{eqnarray}\label{magneticmoment2}
\Pi_{i}&=&\int_{m_{Q}^{2}}^{s_{0}}e^{\frac{-s}{M^{2}}}\rho_{i}(s)ds+e^{\frac{-m_Q^2}{M^{2}}}\Gamma_{i},
\end{eqnarray}
where the explicit expressions for the $\rho_{i}$ and $\Gamma_{i}$ functions are given in \cite{guzel}. In the above relations $M^2$ and $s_{0}$ are the Borel mass square and continuum threshold, respectively.

For calculation of the magnetic moments of the considered baryons,
their residues $\lambda_{B}$ as well as their masses are needed (see
Eq. (\ref{magneticmoment1})). Note that many of the considered
baryons are not discovered yet in the experiments.  The residue is
determined from  analysis of the two point sum rules. For the interpolating current
given in Eq. (\ref{currentguy}), we obtain the following result for
$\lambda_{B}^{2}$:
\begin{eqnarray}\label{resediueguy1}
\lambda_{B}^{2}e^{\frac{-m_{B_{Q}}^{2}}{M^{2}}}=A^2\left[\vphantom{\int_0^{x_2}}\Pi'+\Pi'(q_{1}\longleftrightarrow q_{2})\right],
\end{eqnarray}
where the explicit expression for $\Pi'$ is presented in \cite{guzel}.
The masses of the considered baryons can  be determined from the sum rules. For this aim, one can get the derivative from both side of Eq. (\ref{resediueguy1})
 with respect to $-1/M^2$ and divide the obtained result to the Eq. (\ref{resediueguy1}), i.e.,
\begin{eqnarray}\label{mass}
m_{B_{Q}}^2=\frac{-\frac{d}{d(1/M^2)}\left[\vphantom{\int_0^{x_2}}\Pi'+\Pi'(q_{1}\longleftrightarrow q_{2})\right]}{\left[\vphantom{\int_0^{x_2}}\Pi'+\Pi'(q_{1}\longleftrightarrow q_{2})\right]}.
\end{eqnarray}
\section{Numerical analysis}\label{sec:3}

In this section, we perform numerical analysis for the mass and magnetic moments of the heavy flavored  baryons. Firstly, we present the input parameters used in the analysis of the sum rules: $\uu(1~GeV) = \dd(1~GeV)= -(0.243)^3~GeV^3$, $\ss(1~GeV) = 0.8
\uu(1~GeV)$, $m_0^2(1~GeV) = (0.8\pm0.2)~GeV^2$ \cite{Belyaev},  $\Lambda =
1~GeV$ and $f_{3 \gamma} = - 0.0039~GeV^2$ \cite{Ball}. The value
of the magnetic susceptibility  was
obtained  in various papers as $\chi(1~GeV)=-3.15\pm0.3~GeV^{-2}$  \cite{Ball},  $\chi(1~GeV)=-(2.85\pm0.5)~GeV^{-2}$
\cite{Rohrwild} and   $\chi(1~GeV)=-4.4~GeV^{-2}$\cite{Kogan}.

Before proceeding to the results for the magnetic moments, we calculate the
masses of heavy flavored baryons predicted from mass sum rule.
Obviously, the masses should not depend on the Borel mass parameter
$M^2$ in a complete theory. However, in sum rules method the
operator product expansion (OPE) is truncated and as a result the
dependency of the predictions of physical quantities on the auxiliary parameter
$M^2$ appears. For this reason one should look for a region of $M^2$
such that the predictions for the  physical quantities do not vary with respect to the
Borel mass parameter. This region is the so called the ``working
region`` and within this region the truncation is reasonable and meaningful. The upper
limit of $M^2$ is determined from condition that the continuum and
higher states contributions should be small than the total
dispersion integral. The lower limit is determined by demanding that
in the truncated OPE the condensate term with highest dimension
remains small than sum of all terms, i.e., convergence of OPE should
be under control.

These both conditions conditions for bottom (charmed) baryons
are satisfied when $M^2$ varies in the interval $15~GeV^2<M^2<30~GeV^2$
( $4~GeV^2<M^2<12~GeV^2$). In Figs. \ref{fig:1} -\ref{fig:6}, we
presented the dependence of the mass of the heavy flavored  baryons on $M^2$. From these figures, we see very
good stability with respect to $M^2$.

The sum rule predictions of
the mass of the heavy flavored baryons are presented in Table \ref{tab:2} in
comparison with some theoretical predictions and experimental
results. Note that the masses of the heavy flavored baryons are calculated in
the framework of heavy quark effective theory (HQET) using the QCD
sum rules method in \cite{X.Liu}.
\begin{table}
\caption{Comparison of   mass of the heavy flavored  baryons in  $GeV$ from
present work and other  approaches and with experiment.}
\label{tab:2}       
%
%
\begin{tabular}{p{1.5cm}p{1.5cm}p{1.5cm}p{1.5cm}p{1.5cm}p{1.5cm}p{1.5cm}}
\hline\noalign{\smallskip}
&$m_{\Omega_{b}^{*}}$ & $m_{\Omega_{c}^{*}}$& $m_{\Sigma_{b}^{*}}$ &$m_{\Sigma_{c}^{*}}$  & $m_{\Xi_{b}^{*}}$& $m_{\Xi_{c}^{*}}$ \\
\noalign{\smallskip}\svhline\noalign{\smallskip}
this work&$6.08\pm0.40$&$2.72\pm0.20$ &$5.85\pm0.35$&$2.51\pm0.15 $&$5.97\pm0.40 $&$2.66\pm0.18 $\\
\cite{X.Liu}&$6.063^{+0.083}_{-0.082}$&$2.790^{+0.109}_{-0.105}$
&$5.835^{+0.082}_{-0.077}$&$2.534^{+0.096}_{-0.081}$&$5.929^{+0.088}_{-0.079}$&$2.634^{+0.102}_{-0.094}$\\
\cite{DE}&6.088 &2.768 &5.834&2.518&5.963&2.654\\
\cite{CA}&-&-&5.805&2.495&-&-\\
\cite{RR}&6.090&2.770&5.850&2.520&5.980&2.650\\
\cite{MJ}&-&2.768&-&2.518&-&-\\
\cite{EJ}&6.083&2.760&5.840&-&5.966&-\\
\cite{NM}&6.060&2.752&5.871&2.5388&5.959&2.680\\
Exp\cite{WMY}&-&2.770&5.836&2.520&-&2.645\\
\noalign{\smallskip}\hline\noalign{\smallskip}
\end{tabular}
\end{table}

After determination of the mass as well as residue of the heavy flavored baryons our next task is the calculation of the numerical values of their magnetic moments. For this aim, from sum rules for the magnetic moments it follows that the photon DA's are needed \cite{Ball}. The sum rules for magnetic moments also contain the auxiliary
parameters: Borel parameter $M^2$ and continuum threshold $s_{0}$.
Similar to mass sum rules, the magnetic moments should  also be
independent of these parameters. In the general case, the working region
of $M^2$ and $s_{0}$ for the mass and magnetic moments should be
different. To find the working region for  $M^2$, we proceed as
follows. The upper bound is obtained requiring that the contribution
of higher states and continuum should be less than the ground state
contribution. The lower bound of $M^2$ is determined from condition
that the highest power of $1/M^{2}$ be less than  say $30^0/_{0}$ of
the highest power of $M^{2}$. These two conditions are both
satisfied in the region $15~GeV^2\leq M^{2}\leq30~GeV^2 $ and
$4~GeV^2\leq M^{2}\leq12~GeV^2 $ for baryons containing b and
c-quark, respectively. The working region for the Borel parameter
for mass and magnetic moments practically coincide, but again we
should stress that, this circumstance is accidental.

In Figs. \ref{fig:7}-\ref{fig:16}, we present the dependence of the magnetic moment of
 heavy flavored baryons on $M^2$ at two fixed values of
continuum threshold $s_{0}$. From these figures, we see that the
magnetic moments weakly depend on $s_{0}$. The maximal change of
results is about $10^{~0}/_{0}$ with variation of $s_{0}$. The magnetic
moments also are practically insensitive to the variation of Borel
mass parameter when it varies in the working region. We should also
stress that our results practically don't change considering three
values of $\chi$ which we presented at the beginning of this
section.  Our final results on the magnetic moments of heavy
flavored baryons are presented in Table \ref{tab:3}. For comparison, the
predictions of hyper central model \cite{Patel} are also presented.
The quoted errors in Table 3 are due to the uncertainties in
$m_{0}^2$, variation of $s_{0}$ and $M^2$ as well as errors in the
determination of the input parameters.
\begin{table}
\caption{The    magnetic moments of the heavy flavored baryons in units of nucleon magneton.}
\label{tab:3}      
\begin{tabular}{p{2cm}p{2.4cm}p{3.2cm}}
\hline\noalign{\smallskip}
 & Our results&hyper central model\cite{Patel} \\
\noalign{\smallskip}\svhline\noalign{\smallskip}
$\mu_{\Omega_{b}^{*-}}$ &$-1.40\pm0.35$&$-1.178\div-1.201$ \\
$\mu_{\Omega_{c}^{*0}}$&$-0.62\pm0.18$ &$-0.827\div-0.867$ \\
 $\mu_{\Sigma_{b}^{*-}}$&$-1.50\pm0.36$&$-1.628\div-1.657$\\
 $\mu_{\Sigma_{b}^{*0}}$&$0.50\pm0.15$&$0.778\div0.792$\\
 $\mu_{\Sigma_{b}^{*+}}$&$2.52\pm0.50$&$3.182\div3.239$\\
$\mu_{\Sigma_{c}^{*0}}$&$-0.81\pm0.20$&$-0.826\div-0.850$\\
$\mu_{\Sigma_{c}^{*+}}$&$2.00\pm0.46$&$1.200\div1.256$\\
$\mu_{\Sigma_{c}^{*++}}$&$4.81\pm1.22$&$3.682\div3.844$\\
 $\mu_{\Xi_{b}^{*-}}$&$-1.42\pm0.35$&$-1.048\div-1.098$\\
 $\mu_{\Xi_{b}^{*0}}$&$0.50\pm0.15$&$1.024\div1.042$\\
$\mu_{\Xi_{c}^{*0}}$&$-0.68\pm0.18$&$-0.671\div-0.690$\\
$\mu_{\Xi_{c}^{*+}}$&$1.68\pm0.42$&$1.449\div1.517$\\
\noalign{\smallskip}\hline\noalign{\smallskip}
\end{tabular}
\end{table}

Although the $SU(3)_f$ breaking effects have been
taken into account through a nonzero $s$-quark mass and different strange quark condensate, we predict that $SU(3)_f$ symmetry violation in the magnetic
moments is very small, except the relations $\mu_{\Sigma_{c}^{*+}}=\mu_{\Xi_{c}^{*+}}$ and $\Pi^{\Sigma^{*++}_c} + \Pi^{\Omega^{*0}_c} = 2 \Pi^{\Xi^{*+}_c}$, where the $SU(3)_f$ symmetry violation is large. For the values of the magnetic moments, our results are consistent with the results of \cite{Patel} except for the  $\mu_{\Omega_{b}^{*-}}$,   $\mu_{\Xi_{b}^{*-}}$ and especially for the $\mu_{\Sigma_{c}^{*+}}$, $\mu_{\Xi_{b}^{*0}}$  which we see a big discrepancy between two predictions.

In summary, inspired by recent experimental discovery of the heavy and flavored baryons \cite{Aaltonen1,Abazov,Aaltonen2},
the mass and magnetic moments of these baryons with $J^P=3/2^+$ are calculated within the QCD sum rules. Our results on the masses    are consistent with the experimental data as well as predictions of other approaches. Our results on the masses of the $\Omega_{b}^{*}$,  and $\Xi_{b}^{*}$ can be tested in experiments which will be held in the near future. The predictions on the magnetic moments also can verified in the future experiments.

\begin{acknowledgement}
K. A thanks the organizers of the Crimean conference (2008) for their hospitality. Two of the authors (K. A. and A. O.), would like to thank TUBITAK and TUBA for their partial financial support.
\end{acknowledgement}

\newpage
%
%
%

\end{document}